\documentclass[prb,aps,floatfix,twocolumn,amsmath,amssymb,superscriptaddress]{revtex4-1}
\usepackage{bm}
\usepackage{graphicx}
\usepackage{color}
\usepackage{xcolor}
\usepackage[urlcolor=blue]{hyperref}
\hypersetup{
    colorlinks = true,
    citecolor = {blue},
    linkcolor = {purple},
           }
\setcitestyle{numbers,square}
\allowdisplaybreaks

\begin{document}

\title{Thermal algebraic-decay charge liquid driven by competing short-range Coulomb repulsion}

\author{Ryui Kaneko}
\email{rkaneko@issp.u-tokyo.ac.jp}
\affiliation{%
 Institute for Solid State Physics, University of Tokyo,
 Kashiwa 277-8581, Japan}
\affiliation{%
 International Center for Materials Nanoarchitectonics (WPI-MANA),
 National Institute for Materials Science, Tsukuba 305-0044, Japan}
\author{Yoshihiko Nonomura}
\affiliation{%
 International Center for Materials Nanoarchitectonics (WPI-MANA),
 National Institute for Materials Science, Tsukuba 305-0044, Japan}
\author{Masanori Kohno}
\affiliation{%
 International Center for Materials Nanoarchitectonics (WPI-MANA),
 National Institute for Materials Science, Tsukuba 305-0044, Japan}
\affiliation{%
 Research Center for Functional Materials,
 National Institute for Materials Science, Tsukuba 305-0003, Japan}

\date{\today}

\begin{abstract}
We explore the possibility of
a Berezinskii-Kosterlitz-Thouless-like
critical phase for the charge degrees of freedom in the
intermediate-temperature regime
between the charge-ordered and disordered phases
in two-dimensional systems with
competing short-range Coulomb repulsion.
As the simplest example, we investigate the extended
Hubbard model
with on-site and nearest-neighbor Coulomb interactions
on a triangular lattice at half filling in the atomic
limit by using a classical Monte Carlo method, and find a critical phase,
characterized by algebraic decay of the charge correlation function,
belonging to the universality class of the two-dimensional XY model with a
$\mathbb{Z}_6$ anisotropy. Based on the results, we discuss possible conditions
for the critical phase in materials. 
\end{abstract}


\maketitle


\section{Introduction}

\begin{figure}[b]
\includegraphics[width=0.9\linewidth]{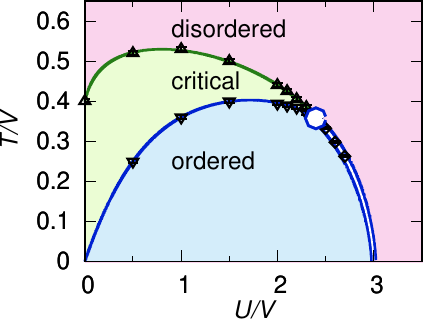}
\caption{Phase diagram of the extended Hubbard model
with on-site and nearest-neighbor Coulomb interactions
in the atomic limit
obtained by Monte Carlo calculations.
The parameters $U(>0)$ and $V(>0)$ denote
the on-site and nearest-neighbor Coulomb repulsion, respectively,
and $T$ indicates temperature.
The first-order transition line (double line) separates
the charge-ordered and disordered phases,
and bifurcates at the critical end point (open
circle) into the boundaries
(single lines)
of the critical phase.}
\label{fig:phase_diag}
\end{figure}

Topological states have recently attracted increasing attention in
search of novel states of matter in materials.
Topological insulators~\cite{hasan2010},
topological superconductors~\cite{sato2017},
quantum Hall systems~\cite{hansson2017},
Kitaev-model systems~\cite{hermanns2017},
symmetry-protected topological
states~\cite{affleck1987,senthil2015},
and
Haldane-gap systems~\cite{haldane1981,kennedy1992}
exhibit features characterized
by topology. The Berezinskii-Kosterlitz-Thouless (BKT) phase is also a
topological phase which exhibits power-law decay of the correlation
function in contrast to a high-temperature disordered phase exhibiting
an exponential decay or a low-temperature long-range ordered
phase~\cite{berezinskii1971,kosterlitz1973}.
Although the BKT phase is known to appear in the Coulomb gas model with
long-range (logarithmic) interactions or in 
planar classical spin
systems in two
spatial dimensions, our motivation is to explore a BKT-like critical insulating
phase for the charge degrees of freedom with short-range Coulomb
repulsion. We consider the following conditions for the critical phase:
(i) two spatial dimensions in favor of topological features, (ii)
competing Coulomb interactions that cause degeneracy for charge
configurations, and (iii) intermediate temperature which is not too low to
stabilize charge order and not too high to result in a
disordered state. As the simplest model, we consider the extended
Hubbard model 
with on-site and nearest-neighbor Coulomb interactions
on a triangular lattice. This model has been investigated
using numerical simulations primarily at zero temperature and
properties of it have been discussed particularly in relation to organic
materials~\cite{hotta2003,
mori2003,
watanabe2006,
udagawa2007,
canocortes2011,
hotta2012,
yoshimi2012,
merino2013,
tocchio2014,
mishmash2015,
gomes2016,
kaneko2017}
and adatoms on semiconductor
surfaces~\cite{tosatti1974,carpinelli1996,santoro1999,hansmann2013,adler2018}.
On the other hand, nonzero-temperature properties of the model have
drawn little attention so far. Possible critical behavior as well as the
finite-temperature phase diagram has never been investigated before.
In the present paper we focus attention on the half filling in
the insulating atomic limit by neglecting the hopping term.
This simplification removes complexity arising from the fermionic
degrees of freedom and allows us to apply a numerically exact classical
Monte Carlo method to large systems at nonzero temperatures, and to
uncover the presence of the charge BKT-like critical phase in the
strongly correlated region of the two-dimensional Hubbard model, as
shown in Fig.~\ref{fig:phase_diag}.

The present paper is organized as follows: In
Sec.~\ref{sec:model_and_method}, we introduce the extended Hubbard model
in the atomic limit on a triangular lattice. In
Sec.~\ref{sec:results}, we present details of
the finite-temperature phase diagram (Fig.~\ref{fig:phase_diag}),
and show charge properties in each phase. We offer the numerical
evidence of the BKT-like critical phase for the charge degrees of
freedom and discuss its nature. Moreover, we explain how 
the spin degrees of freedom modify the phase diagram
in the extended Hubbard model by comparing the
phase diagram with that of the Blume-Capel model,
which corresponds to an effective model only for charge degrees of
freedom.
In Sec.~\ref{sec:discussions}, we discuss possible conditions for the
critical phase realized in materials. Finally, in Sec.~\ref{sec:summary}, we draw
our conclusions.

\section{Model and method}
\label{sec:model_and_method}

\subsection{Extended Hubbard model}

We consider the extended Hubbard model on a triangular lattice
in the atomic limit. The Hamiltonian reads
\begin{equation}
H_{\rm EH} = U\sum_{i} n_{i\uparrow} n_{i\downarrow}
  + V\sum_{\langle i, j\rangle} n_i n_j
  - \mu \sum_{i} n_i,
\end{equation}
where
$U$ ($V$) denotes
the strength of
on-site (nearest-neighbor) Coulomb interaction,
and $\mu$ indicates chemical potential.
Here
$n_i=n_{i\uparrow} + n_{i\downarrow}$,
$n_{i\sigma}\in \{0,1\}$ denotes the number of
electrons with spins $\sigma=\uparrow,\downarrow$ at site $i$,
and
$\langle i, j\rangle$ means that sites $i$ and $j$
are nearest neighbors.
The electron density is given by
\begin{equation}
 \rho = \frac{1}{N_{\rm s}} \sum_{i} \langle n_i \rangle,
\end{equation}
where $N_{\rm s}$ and $\langle \cdots \rangle$ denote
the system size and thermal averaging, respectively.
We consider the half-filled case ($\rho=1$,
$\mu=U/2+zV$ with
the coordination number $z=6$ on a triangular lattice).
Hereafter, we focus on the repulsive
Coulomb interactions ($U,V>0$).

\subsection{Classical Monte Carlo method}
\label{subsec:model_and_method_MC}

We apply the classical Monte Carlo method with
the local-update Metropolis algorithm
to investigate the phase diagrams for the extended Hubbard model.
A similar approach has been applied to
the doped systems on
a square lattice~\cite{misawa2006,pawlowski2006,kapcia2017},
the half-filled ones with long-range Coulomb interactions
on a cubic lattice~\cite{pramudya2011},
and the quarter-filled ones on a triangular lattice~\cite{yoshida2014,mahmoudian2015}.
We perform the grand-canonical Monte Carlo simulation containing
insertion and removal of electrons as well as moving of them~\cite{pawlowski2006}.
We also use the exchange Monte Carlo method~\cite{hukushima1996}
to perform the simulation at low temperatures efficiently.
We typically use $200$ or $400$ replicas distributed evenly
in the temperature region $0<T/V\le 1$.
We adopt the periodic boundary condition, and
consider the system size $N_{\rm s}=L^2$ with
$L=12$, $24$, $48$, $96$, and $192$.
We typically perform $10^5$ Monte Carlo steps for sampling after discarding $10^4$
Monte Carlo steps for thermalization.
We perform $10$ independent runs starting from different random initial conditions
to estimate statistical errors.
Hereafter, we will set the Boltzmann factor $k_{\rm B}=1$,
and choose $V$ as the unit of energy in the extended Hubbard model.

\section{Results}
\label{sec:results}

\subsection{Ground states}
\label{subsec:ground_states}

\begin{figure}[t]
\includegraphics[width=0.9\linewidth]{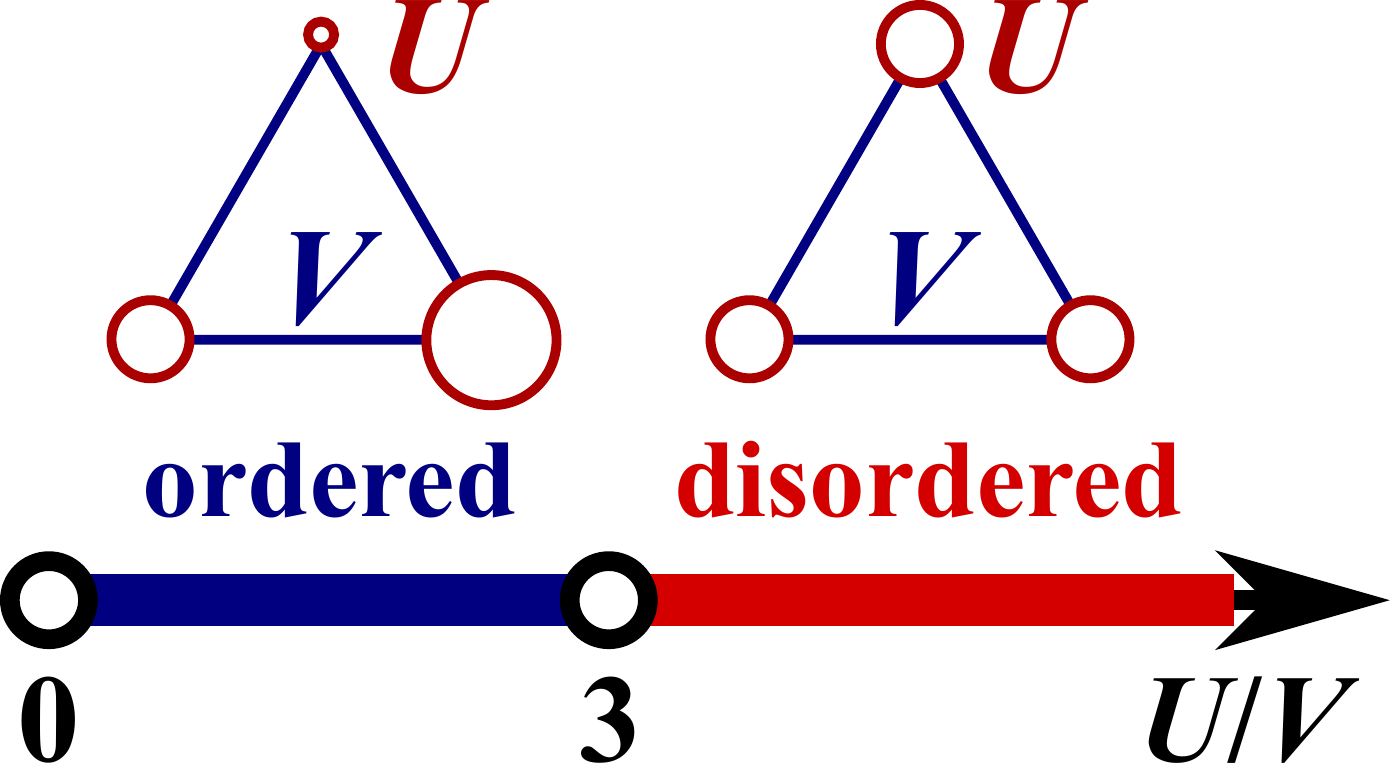}%
\caption{Ground-state phase diagram for $U,V>0$. Large, middle, and small circles
on triangles represent
the sites with $n_i=2$, $n_i=1$, and $n_i=0$,
respectively.}
\label{fig:GSPD}
\end{figure}

Before discussing the phases at nonzero temperatures
of the extended Hubbard model, let us discuss the ground states.
When the on-site Coulomb interaction $U$ is much larger than $V$,
the double occupancy of electrons is prohibited
and the ground state is charge uniform,
where all sites are singly occupied.
On the other hand, when the nearest-neighbor Coulomb interaction $V$
is much larger than $U$,
empty sites reduce the energy loss in the $V n_i n_j$ term.
As a result, the ground state becomes the $012$-type
charge-ordered state
where three-sublattice sites
have $n_i=0$, $1$, and $2$ (see the left side of Fig.~\ref{fig:GSPD}).
Note that
the same charge order configuration is found in the presence of nonzero
hopping~\cite{santoro1999,hansmann2013,hansmann2016}.
The energy per site for the former state is $E/N_{\rm s}=3V$
while that for the latter state is $E/N_{\rm s}=2V+U/3$,
and hence, the level crossing occurs at
$U/V=3$
(see a schematic ground-state phase diagram in Fig.~\ref{fig:GSPD}).
Hereafter, we mainly focus on the realistic parameter region
$U\gtrsim V$.

Note that the classical ground state has macroscopic degeneracy exactly at $U=0$.
Each local triangle can have one of the $\{012, 002, 022\}$ charge configurations
with the conserved total number of electrons.
Besides, in the presence of very strong attractive on-site Coulomb interaction
($U\rightarrow -\infty$), electrons bind together, and singly occupied sites disappear.
The achievable number of electrons per site is only $n_i=0$ or $2$,
and the system effectively becomes the antiferromagnetic Ising model
on a triangular lattice~\cite{wannier1950}.
The correlation function of the Ising order parameter ($n_i-1$ in this case)
shows the power-law decay $\sim1/r^{\eta}$
as a function of distance $r$ with the critical exponent $\eta=1/2$
at zero temperature~\cite{stephenson1970,alexander1980}.

\subsection{Phase diagram at nonzero temperatures}

Reflecting the charge order in the ground state, the ordered phase
survives at nonzero temperatures for $U/V<3$
(Fig.~\ref{fig:phase_diag}). The first-order
phase transition occurs from the ordered phase to the disordered phase
along the transition line for $U_{\rm c}/V<U/V<3$. The first-order
transition line bifurcates at the critical end point
$(U_{\rm c}/V,T_{\rm c}/V)=(2.45(5), 0.36(1))$
into two boundaries which enclose a
critical phase. The boundaries are determined from the behavior of the
charge correlation function
which will be explained in Sec.~\ref{subsec:exponent}.

\begin{figure}[t]
\includegraphics[width=0.9\linewidth]{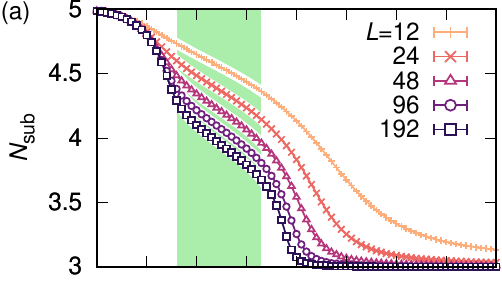}\\
\includegraphics[width=0.9\linewidth]{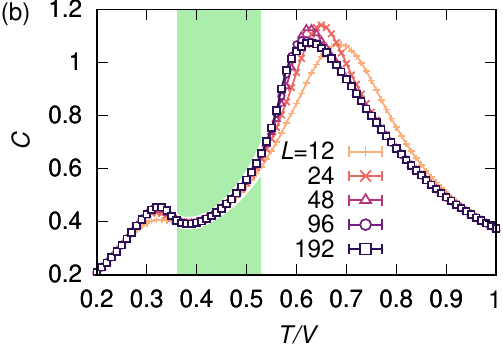}
\caption{Temperature dependence of
(a) the sublattice charge order parameter
and
(b) the specific heat at $U/V=1$
for $L=12$ (crosses), $24$ (X marks), $48$ (triangles),
$96$ (circles), and $192$ (squares).
The shaded area corresponds to the charge critical phase.
The statistical errors are smaller than the size of symbols.}
\label{fig:spec_heat}
\end{figure}

To get insight into the charge critical phase,
we show temperature dependence of the sublattice charge order parameter
\begin{equation}
\label{eq:sub_CO}
 N_{\rm sub} = \langle n_A^2 + n_B^2 + n_C^2 \rangle
\end{equation}
along the $U/V=1$ line in Fig.~\ref{fig:spec_heat}(a).
Here the electron density for each sublattice
$n_{\alpha}$ ($\alpha=A,B,C$) on a triangular lattice is defined by
\begin{equation}
\label{eq:sublattice_el_density}
 n_{\alpha} = \frac{3}{N_{\rm s}}\sum_{i\in \alpha} n_i.
\end{equation}
At low temperatures, the $012$-type charge order develops,
and $N_{\rm sub}$ approaches $5=0^2+1^2+2^2$.
At high temperatures,
charge becomes uniform, each site contains one electron,
and $N_{\rm sub}$ approaches $3=1^2+1^2+1^2$.
By contrast, in the intermediate (shaded) region
between the charge-ordered and disordered phases
in Fig.~\ref{fig:spec_heat}(a),
$N_{\rm sub}$ shows a large size dependence,
and seems to approach $3$ extremely slowly.
This fact suggests the presence of an intermediate region,
where the charge correlation shows nearly critical behavior.

Just above and below the intermediate phase,
the specific heat defined by
\begin{equation}
 C = \frac{\langle H_{\rm EH}^2 \rangle - \langle H_{\rm EH} \rangle^2}{T^2}
\end{equation}
shows broad two peaks, as shown in Fig.~\ref{fig:spec_heat}(b).
Although the position of the lower-temperature peak at $T/V\simeq 0.3$
is almost independent of sizes, that of the higher-temperature
peak at $T/V\simeq 0.6$ approaches lower temperatures
as the system size increases. Nevertheless, the peak values
do not change significantly, which suggests
that this behavior is 
a crossover or a BKT-like transition.

\subsection{Density of doubly occupied sites}
\label{subsec:def_rho_2}

\begin{figure}[t]
\includegraphics[width=0.9\linewidth]{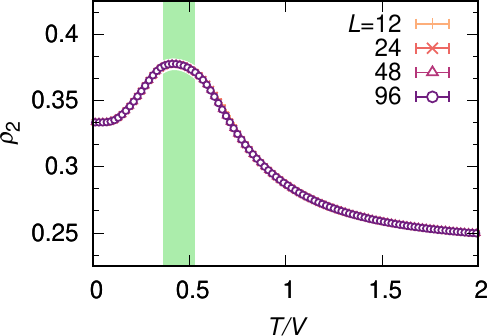}
\caption{Density of
doubly occupied sites as a function of temperature at $U/V=1$
for $L=12$ (crosses), $24$ (X marks), $48$ (triangles),
and $96$ (circles).
The shaded area corresponds to the charge critical phase.
There exists almost no size dependence within the size of symbols.}
\label{fig:dbl_occ}
\end{figure}

We also calculate temperature dependence of the density of
doubly occupied sites defined by
\begin{equation}
 \rho_2 = \frac{1}{N_{\rm s}} \sum_{i}
 \langle n_{i\uparrow} n_{i\downarrow} \rangle,
\end{equation}
as shown in Fig.~\ref{fig:dbl_occ}.
At half filling, the density of empty sites $\rho_0$ is equal to
$\rho_2$ since empty and doubly occupied sites are simultaneously
created when an electron
with a spin $\sigma$ moves from a singly occupied site to another
singly occupied site.
Thus, the density of singly occupied sites is given as
$\rho_1 = 1 - 2\rho_2$.
In the $012$-type charge-ordered state at zero temperature,
$\rho_0=\rho_1=\rho_2=1/3$.
In the high-temperature limit, since all the interactions can be
neglected, $\rho_0 = \rho_{\uparrow} = \rho_{\downarrow} = \rho_2 = 1/4$,
where the density of singly occupied sites
by a spin $\sigma$ electron is denoted by $\rho_{\sigma}$.
Here $\rho_{\uparrow} = \rho_{\downarrow} = \rho_{1}/2$ without
magnetic field.
Remarkably, $\rho_2$ is not monotonic as a function of $T$,
and becomes a maximum in the critical phase.

\begin{figure}[t]
\includegraphics[width=0.625\linewidth]{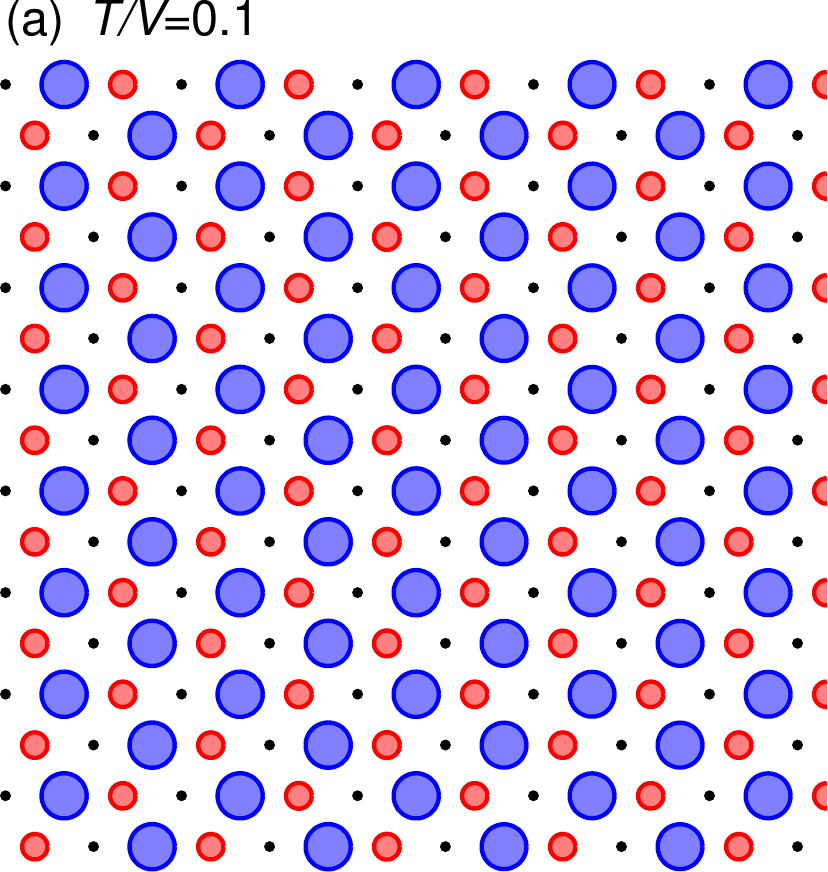}\\
\bigskip
\includegraphics[width=0.625\linewidth]{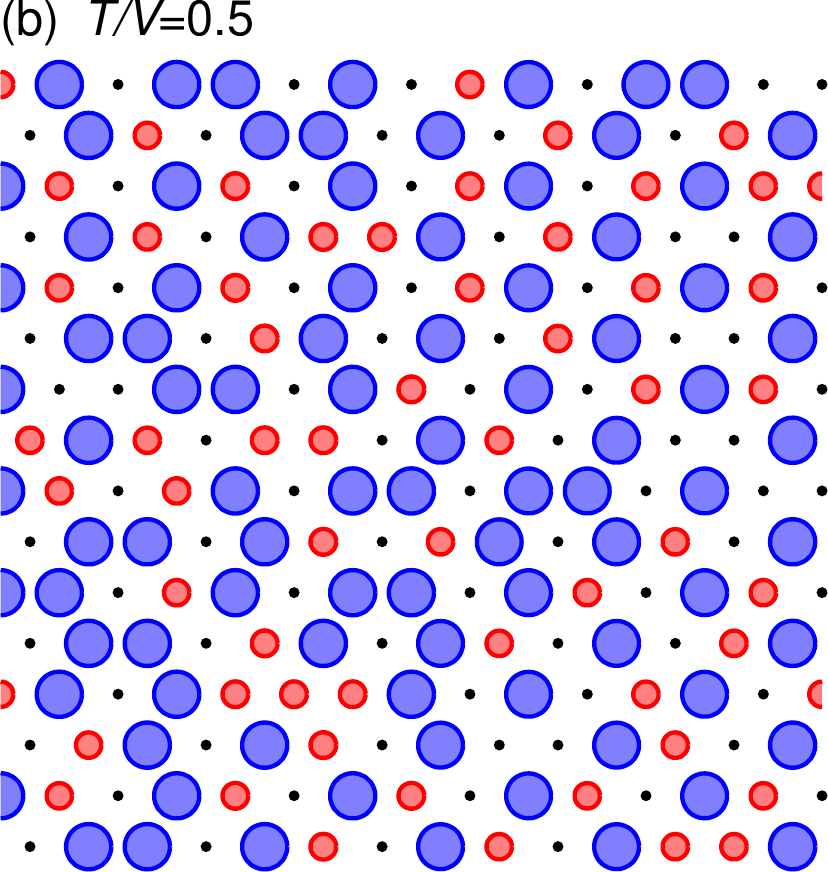}\\
\bigskip
\includegraphics[width=0.625\linewidth]{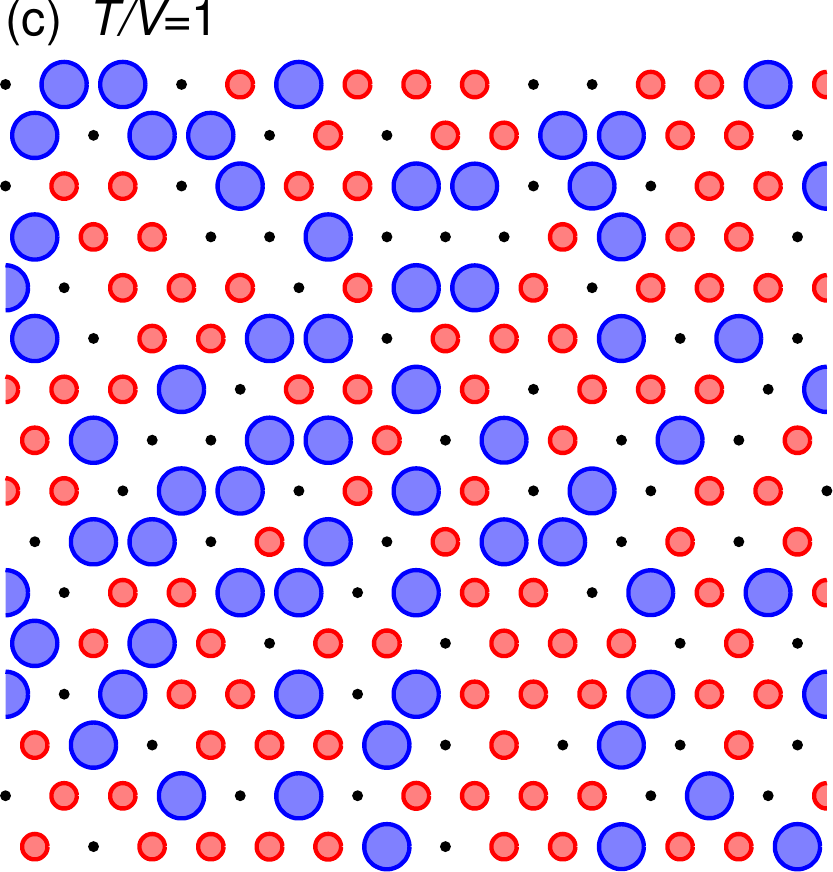}
\caption{Snapshots of charge configuration for $U/V=1$ at
(a) $T/V=0.1$,
(b) $T/V=0.5$,
and
(c) $T/V=1$.
The system size is $N_{\rm s}=24\times 24$.
Large, middle, and small circles represent the sites with
$n_i=2$, $n_i=1$, and $n_i=0$, respectively.}
\label{fig:snapshot}
\end{figure}

For further understanding of the behavior of charge,
we show the snapshots of charge configuration in the real space
in Fig.~\ref{fig:snapshot}.
As the temperature is raised, the $012$-type
charge order [see Fig.~\ref{fig:snapshot}(a)] in the low-temperature
regime is partially broken by thermal
fluctuations, and $002$ or $022$ triangles
[see Fig.~\ref{fig:snapshot}(b)] are induced in the
intermediate-temperature regime.
Thus, $\rho_1$ decreases, while
$\rho_2[=(1-\rho_1)/2]$
increases (see Fig.~\ref{fig:dbl_occ}).
On the other hand, in the high-temperature regime,
the charge distribution becomes more random
[see Fig.~\ref{fig:snapshot}(c)], and $\rho_1$ ($\rho_2$)
increases (decreases)
(see Fig.~\ref{fig:dbl_occ}).

While the charge particles themselves interact with each other
with the logarithmic potential in the two-dimensional Coulomb
gas model, defects surrounded by ordered domains correspond
to the vortices and antivortices in the present critical phase, and
the topological nature of it should be
equivalent to that found 
in Ising antiferromagnets on a triangular
lattice~\cite{landau1983}.

\subsection{Emergent $\rm U(1)$ symmetry}

To see the $\rm U(1)$ symmetry induced by thermal fluctuation in the
critical phase, we define the radius variable $R$ and azimuth
variable $\theta$ as
\begin{eqnarray}
 N_x+iN_y &=& R e^{i\theta},
\\
 R &=& \sqrt{N_x^2 + N_y^2},
\\
 \mbox{and}\quad
 \theta &=& \arctan \frac{N_y}{N_x},
\end{eqnarray}
with
\begin{eqnarray}
 N_x &=& \frac{2n_A-n_B-n_C}{\sqrt{6}}
\\
 \mbox{and}\quad
 N_y &=& \frac{n_B-n_C}{\sqrt{2}}
\end{eqnarray}
by using the electron density for each sublattice
$n_{\alpha}$ ($\alpha=A,B,C$)
in Eq.~(\ref{eq:sublattice_el_density}).
The equivalent variables for the spin degrees of freedom have been used
in previous studies, especially, in Ising antiferromagnets
on a triangular lattice~\cite{landau1983,takayama1983,%
fujiki1986,%
isakov2003,%
ishizuka2013,%
nishino2016,%
wang2017}.

\begin{figure}[t]
\includegraphics[width=0.9\linewidth]{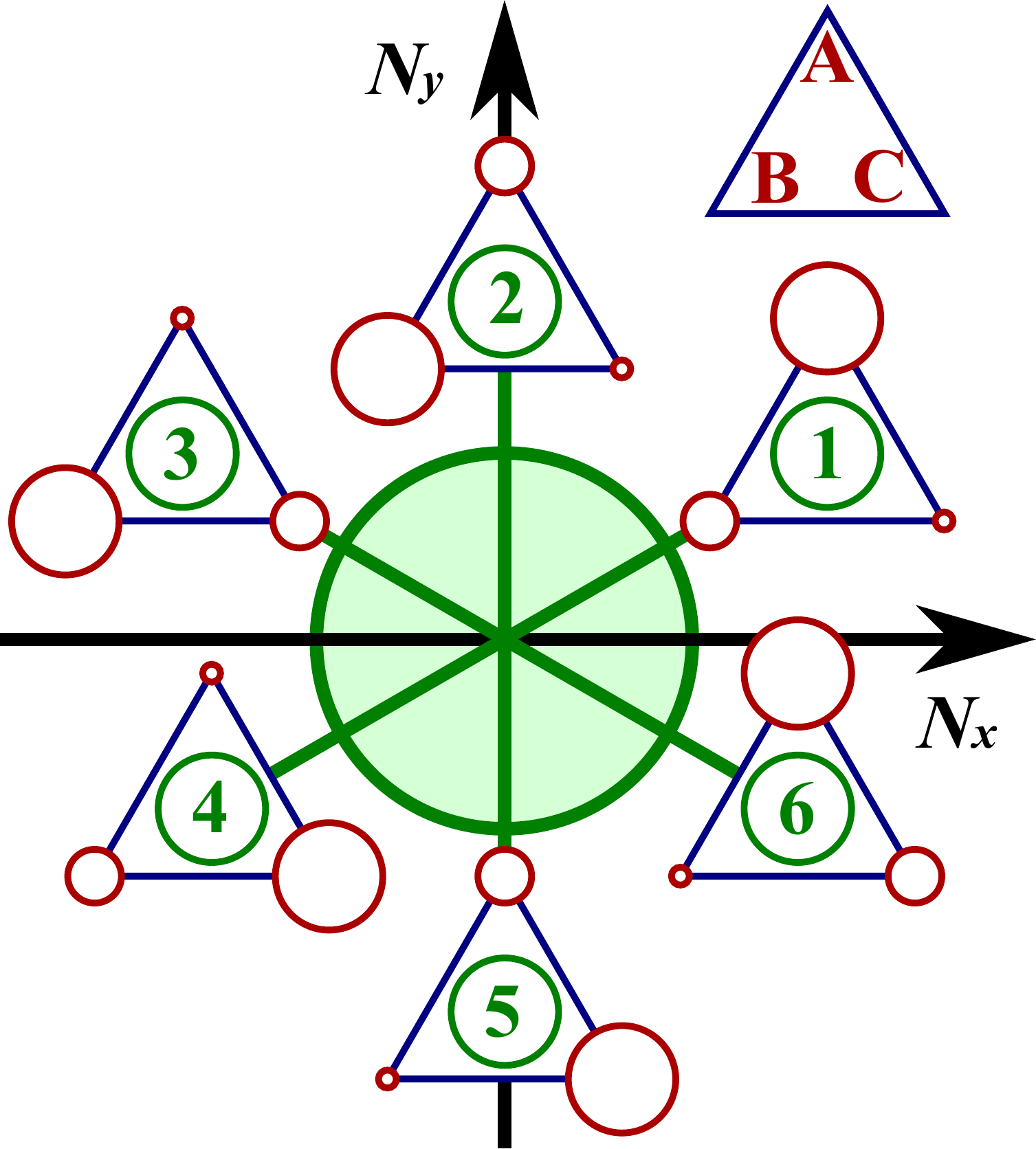}
\caption{Radius and azimuth 
variables for each charge-ordered state.
The state number in each triangle corresponds to that in Table~\ref{tab:azimuth}.
Large, middle, and small circles represent the sites with
$n_i=2$, $n_i=1$, and $n_i=0$, respectively.}
\label{fig:six_state}
\end{figure}

The charge-ordered states are sixfold degenerate,
and we can assign the radius and azimuth variables 
to each state,
as shown in Table~\ref{tab:azimuth}
and Fig.~\ref{fig:six_state}.
The squared radius is proportional
to the charge structure factor at the ordering wave vector
$\bm{Q}=(4\pi/3,0), (2\pi/3,2\pi/\sqrt{3})$
on the hexagonal Brillouin zone:
\begin{eqnarray}
\label{eq:Nxy_prop_to_Nq}
 \langle R^2 \rangle
 &=& \frac{1}{3}\langle (n_A-n_B)^2 + (n_B-n_C)^2 +(n_C-n_A)^2 \rangle
\nonumber
\\
 &=& \frac{6N(\bm{Q})}{N_{\rm s}},
\end{eqnarray}
where the charge structure factor $N(\bm{q})$ is defined by
\begin{equation}
 N(\bm{q}) = \frac{1}{N_{\rm s}} \sum_{i,j}
 (\langle n_i n_j \rangle - \rho^2)
 e^{i\bm{q}\cdot(\bm{r}_i-\bm{r}_j)}.
\end{equation}
When the long-range order appears,
the azimuths satisfy
$\langle \cos 6\theta \rangle = -1$, and the squared radius
$\langle R^2 \rangle$ gives a nonzero value
in the thermodynamic limit.

\begin{table}[t]
\caption{Radius and azimuth variables
for the sixfold-degenerate ground states
(see Fig.~\ref{fig:six_state}) of the charge degrees of freedom.}
\label{tab:azimuth}
{\renewcommand{\arraystretch}{1.5}
\setlength{\tabcolsep}{3.5pt}
\begin{tabular}{c||cccccc}
\hline
State \# & $1$ & $2$ & $3$ & $4$ & $5$ & $6$ \\
\hline
\hline
$n_A$ & $2$ & $1$ & $0$ & $0$ & $1$ & $2$ \\
$n_B$ & $1$ & $2$ & $2$ & $1$ & $0$ & $0$ \\
$n_C$ & $0$ & $0$ & $1$ & $2$ & $2$ & $1$ \\
\hline
$N_x$ & $+3/\sqrt{6}$ & $0$ & $-3/\sqrt{6}$ & $-3/\sqrt{6}$ & $0$ & $+3/\sqrt{6}$ \\
$N_y$ & $+1/\sqrt{2}$ & $+\sqrt{2}$ & $+1/\sqrt{2}$ & $-1/\sqrt{2}$ & $-\sqrt{2}$ & $-1/\sqrt{2}$ \\
\hline
$R$ & $\sqrt{2}$ & $\sqrt{2}$ & $\sqrt{2}$ & $\sqrt{2}$ & $\sqrt{2}$ & $\sqrt{2}$ \\
$\theta$ & $+\pi/6$ & $+\pi/2$ & $+5\pi/6$ & $-5\pi/6$ & $-\pi/2$ & $-\pi/6$ \\
\hline
\end{tabular}
}
\end{table}

\begin{figure}[t]
\includegraphics[width=0.9\linewidth]{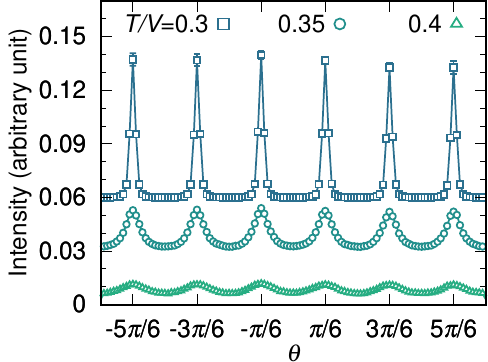}
\caption{Histogram of the azimuth variable
$\theta$ at $U/V=1$.
The system size is $N_{\rm s}=48\times 48$.
The intensity is shifted upward by $0.03$
with decreasing temperature
for clear visualization.}
\label{fig:hist_theta}
\end{figure}

The ground state is characterized by the radius $R=\sqrt{2}$
and the azimuths $\theta=\pi/6+n\pi/3$ ($n\in\mathbb{Z}$)
(see Table~\ref{tab:azimuth}).
As shown in Fig.~\ref{fig:hist_theta},
these azimuths
appear more frequently at sufficiently low temperatures,
and the six peaks develop in the histogram,
corresponding to the
charge order
of the ground state.
As temperature increases, the peaks broaden
and become indiscernible.

\begin{figure}[t]
\includegraphics[width=0.75\linewidth]{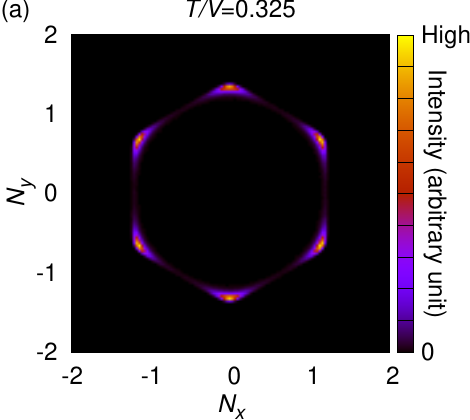}\\
\bigskip
\includegraphics[width=0.75\linewidth]{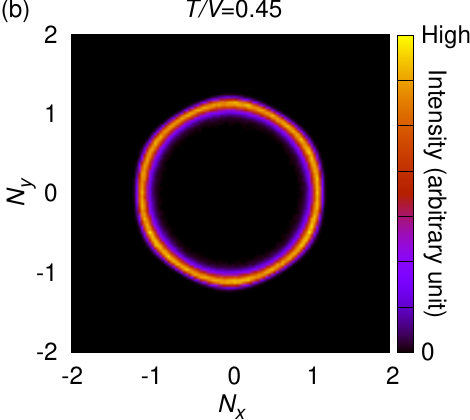}\\
\bigskip
\includegraphics[width=0.75\linewidth]{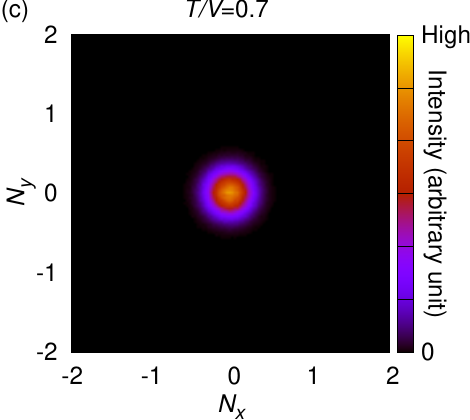}
\caption{Histogram of the 
two-component vector $(N_x,N_y)$ for $U/V=1$
for the system size $N_{\rm s}=48\times 48$ at
(a) $T/V=0.325$,
(b) $T/V=0.45$,
and
(c) $T/V=0.7$.}
\label{fig:hist_num}
\end{figure}

The behavior of charge melting is visualized more clearly
in the histogram of the 
two-component vector $(N_x,N_y)$.
At a low temperature ($T/V=0.325$), the six sharp peaks appear at
$R\simeq\sqrt{2}$ and $\theta=\pi/6+n\pi/3$ ($n\in\mathbb{Z}$)
[see Fig.~\ref{fig:hist_num}(a)].
At an intermediate temperature ($T/V=0.45$),
a ring structure develops instead of the discrete peaks.
This behavior suggests the melting of charge order
and the emergence of $\rm U(1)$ symmetry 
by thermal fluctuations
[see Fig.~\ref{fig:hist_num}(b)],
and the radius $R$ shrinks as temperature increases.
At a high temperature ($T/V=0.7$), $R$ becomes nearly zero,
which suggests absence of the long-range order
[see Fig.~\ref{fig:hist_num}(c)].
Note that
$\langle R^2 \rangle$ is scaled with $L^{-\eta}$ ($\eta>0$)
for large $L$ in the intermediate-temperature region
(see Sec.\ref{subsec:exponent}).

Such an emergent $\rm U(1)$ symmetry has been discussed in
the two-dimensional $p$-state clock model,
which is a discrete version of the XY model~\cite{jose1977}.
For $p\rightarrow\infty$, the system is identical to the XY model,
and the BKT phase appears at nonzero temperatures.
For a finite but large enough $p$,
thermal fluctuations effectively smear out discreteness of the order
parameter, and still induce the BKT phase.
At the same time,
long-range order remains at finite temperatures.
The renormalization-group analysis~\cite{jose1977}
suggests that two BKT transitions appear
for $p>p_{\rm c}$ with $4<p_{\rm c}<5$.
The present system would correspond to the case of $p=6$,
which is above the critical value $p_{\rm c}$.

\subsection{Critical exponent}
\label{subsec:exponent}

Here we investigate the critical behavior of the BKT-like transitions
to show that
the intermediate phase can be identified with that of
the two-dimensional XY model with a $\mathbb{Z}_6$ anisotropy.

\begin{figure}[t]
\includegraphics[width=0.9\linewidth]{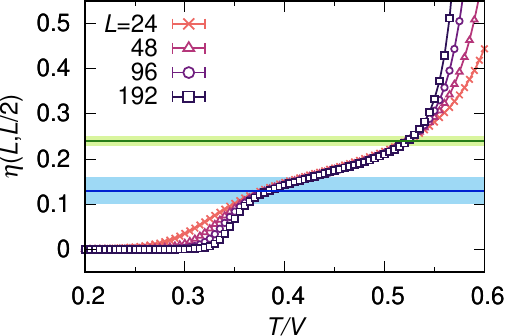}
\caption{Temperature dependence of $\eta(L,L/2)$ at $U/V=1$
for $L=12$ (crosses), $24$ (X marks), $48$ (triangles),
$96$ (circles), and $192$ (squares).
The horizontal lines indicate the boundaries of the critical phase
with error bars (shaded areas).}
\label{fig:exponent_eta}
\end{figure}

In the critical region, the charge correlation function shows
algebraic decay as a function of distance $r$
with the critical exponent $\eta$:
\begin{equation}
 \langle n_0 n_r \rangle - \rho^2 \sim r^{-\eta}.
\end{equation}
Then, 
the peak value of charge structure factor is scaled as
\begin{eqnarray}
\label{eq:Nq_prop_to_Lpower}
 N(\bm{Q})
 &=& \sum_{r} (\langle n_0 n_r \rangle - \rho^2) e^{i\bm{Q}\cdot\bm{r}}
\nonumber
\\
 &\sim& \int_{\Lambda}^{L} d^2r \, r^{-\eta}
  \sim  L^{2-\eta},
\end{eqnarray}
where $\Lambda$ is a cutoff,
and 
the squared radius [Eq.~(\ref{eq:Nxy_prop_to_Nq})]
shows the power-law decay as
\begin{equation}
\label{eq:R2_prop_to_Lpower}
 \langle R^2 (L) \rangle
 \sim L^{-\eta}.
\end{equation}
The critical exponent $\eta$ can be estimated~\cite{miyashita1991} by
$\eta=\lim_{L_1,L_2\rightarrow\infty} \eta (L_1,L_2)$ with
\begin{equation}
\label{eq:def_eta}
 \eta (L_1,L_2) =
  - \frac{\ln[\langle R^2 (L_2) \rangle /\langle R^2 (L_1) \rangle ]}
         {\ln(L_2/L_1)}.
\end{equation}

As shown in Fig.~\ref{fig:exponent_eta},
$\eta$ approaches zero at low temperatures
because $\langle R^2\rangle$ converges to a nonzero value
[see Eqs.~(\ref{eq:Nxy_prop_to_Nq}) and (\ref{eq:R2_prop_to_Lpower})].
On the other hand, $\eta=2$ holds at high temperatures
because there is no long-range order
and $N(\bm{Q})$ converges to a constant value
[see Eq.~(\ref{eq:Nq_prop_to_Lpower})].
By contrast, in the intermediate-temperature regime, $\eta(L,L/2)$ for
various $L$ falls on a single curve.
At $U/V=1$,
the exponents $\eta$ at the lower and upper
phase boundaries are estimated as
$\eta_{\rm low}=0.13(3)$ and
$\eta_{\rm high}=0.24(1)$, respectively.
These values are consistent with
the critical exponents $\eta_{\rm low}=1/9$
and $\eta_{\rm high}=1/4$ in the two-dimensional XY model
with a $\mathbb{Z}_6$ anisotropy (the six-state clock model).
We have numerically confirmed that this result holds true for $0<U<U_{\rm c}$.

We identify the boundaries of the critical phase as the temperatures
where $\eta$ obtained in the largest system sizes coincides with
$\eta_{\rm low}=1/9$ and $\eta_{\rm high}=1/4$,
and the phase diagram drawn with this criterion is given
in Fig.~\ref{fig:phase_diag}.

Note that the equivalent criticality has been suggested in
the dislocation-mediated melting
in the triangular system~\cite{nelson1979},
the $J_1$-$J_2$ triangular Ising
antiferromagnet~\cite{landau1983,takayama1983,miyashita1991,sato2013},
the triangular Heisenberg model with a single-ion
anisotropy~\cite{damle2015,heidarian2015},
the Coulomb crystals in trapped ions~\cite{podolsky2016},
and
the triangular Blume-Capel model~\cite{collins1988,ballou1991,grigelionis1994,
zukovic2012,zukovic2013a,zukovic2013b,ibenskas2014},
which will be investigated further below.
In contrast to the Coulomb gas model on a triangular
lattice~\cite{lee1991},
where the BKT transition line hits the first-order transition line
slightly below the critical end point,
the BKT-like transition lines seem to terminate
at a single critical point
for all the systems with short-range interactions mentioned above.
In general, first-order, second-order, and BKT transitions are allowed
in the $\mathbb{Z}_p$ model ($p\ge 5$), which includes additional interaction terms
with relative angles $2n\pi/p$ ($n=2,3,\dots,p-2$) in the $p$-state clock
model~\cite{cardy1980,alcaraz1980,alcaraz1981}.
The critical bifurcation point at nonzero temperature
(see Fig.~\ref{fig:phase_diag}) is inferred
to be the Fateev-Zamolodchikov point~\cite{fateev1982},
where the first-order and BKT transition lines
terminate~\cite{alcaraz1987,rouidi1992,dasilva2014}.

\subsection{Relationship with the effective model without spin degrees of freedom}
\label{subsec:relation_eff_model}

For further clarification of the peculiarity of the critical phase,
let us consider the effects of the spin degrees of freedom
by comparing the above results with those for the model
without spin degrees of freedom. 

\begin{table}[t]
\caption{Degeneracy of $n_i$ in the extended Hubbard model
and that of $S_i$ in the Blume-Capel model.}
\label{tab:degeneracy}
\begin{tabular}{c||c|@{~~}c@{~~}|@{~~}c@{~~}|c}
\hline
 $n_i$ & $0$ & $\uparrow$ & $\downarrow$ & $2$ \\
\hline
 Degeneracy of $n_i$ & $1$ & $1$ & $1$ & $1$ \\
\hline
\hline
 $S_i=n_i-1$ & $-1$ & \multicolumn{2}{c@{~~}|}{$0$} & $+1$ \\
\hline
 Degeneracy of $S_i$ & $1$ & \multicolumn{2}{c@{~~}|}{$2$} & $1$ \\
\hline
\end{tabular}
\end{table}

The extended Hubbard model in the atomic limit can be mapped to the classical
$S=1$ Ising model, namely,
the Blume-Capel model~\cite{blume1971,jedrzejewski1994,pawlowski2006}
with temperature-dependent interactions as follows.
By substituting the electron density $n_i$ with
the $S=1$ Ising variable as $S_i=n_i-1$,
the partition function of the extended Hubbard model is transformed
into that of the Blume-Capel model:
\begin{eqnarray}
 Z &=& \sum_{\{n_i\}} e^{-\beta H_{\rm EH}(n_i)}
  = \sum_{\{S_i\}} e^{-\beta H_{\rm EH}(S_i)} \prod_{i} g(S_i)
\nonumber
\\
  &\propto& \sum_{\{S_i\}} e^{-\beta H_{\rm BC}(S_i)}
\end{eqnarray}
with the entropy factor $g(S_i)$ taken into account
to compensate the twofold spin degeneracy in each singly occupied site
in the extended Hubbard model (see Table~\ref{tab:degeneracy}).
Here $g(S_i)$ can be rewritten as follows:
\begin{equation}
 g(S_i) = \delta_{S_i,-1}+2\delta_{S_i,0}+\delta_{S_i,1} =
 e^{(1-S_i^2) \ln 2}.
\end{equation}
Finally, we obtain the Hamiltonian
\begin{equation}
\label{eq:BC_model}
 H_{\rm BC} = \Delta\sum_{i} S_i^2 + J\sum_{\langle i, j\rangle} S_i S_j
  - h \sum_{i} S_i
\end{equation}
with
\begin{eqnarray}
\label{eq:Delta_eq_halfU_plus_T}
 \Delta &=& \frac{U}{2} + T\ln 2,
\\
 J &=& V,
\\
 \mbox{and}\quad
 h &=& \mu - \frac{U}{2} - zV.
\end{eqnarray}
Here $\Delta$, $J$, and $z$ denote
the strength of single-ion anisotropy,
the strength of nearest-neighbor interaction,
and the coordination number, respectively.
The external magnetic field $h$
takes the place of chemical potential $\mu$ in the extended Hubbard model.
The contribution of spin entropy, whose coefficient is proportional to $T$,
is now absorbed in the single-ion anisotropy $\Delta$.

Hereafter, we consider the Blume-Capel model with
the temperature-independent $\Delta$ at $h=0$,
and compare the properties of this model with those of the extended
Hubbard model at half filling.
Note that 
the extended Hubbard model
has
degeneracy for the spin degrees of freedom
which is absent in the Blume-Capel model,
and therefore,
their finite-temperature phase
diagrams are different.
The presence or absence of the BKT-like phase can be identified only by
calculating the correlation function directly in each model.
However,
the extended Hubbard model and the Blume-Capel model with $\Delta=U/2$
have essentially the same ground state
[see Eq.~(\ref{eq:Delta_eq_halfU_plus_T})].

We use the canonical Monte Carlo method~\cite{zeng1997}
in the Blume-Capel model
with the conditions similar to those in the extended Hubbard model
(see Sec.~\ref{subsec:model_and_method_MC}).
At sufficiently low temperatures,
thermal averaging of the Blume-Capel model
($\langle\cdots\rangle_{\rm BC}$)
is essentially the same as that of the extended Hubbard model
($\langle\cdots\rangle$).
Consequently,
the squared magnetic moment in the Blume-Capel model,
\begin{equation}
 M = \frac{1}{N_{\rm s}}\sum_{i}\langle S_i^2\rangle_{\rm BC},
\end{equation}
is related to the density of doubly occupied sites
$\rho_2$ in the extended Hubbard model by $M=2\rho_2$ at low temperatures
since $\rho_0=\rho_2$ (see Sec.~\ref{subsec:def_rho_2}).
Instead of the charge structure factor in the extended Hubbard model,
we calculate the spin structure factor defined by 
\begin{equation}
 S(\bm{q}) = \frac{1}{N_{\rm s}} \sum_{i,j}
 \langle S_i S_j \rangle_{\rm BC} ~
 e^{i\bm{q}\cdot(\bm{r}_i-\bm{r}_j)}
\end{equation}
in the Blume-Capel model. 
The phase boundaries are determined by the critical exponent $\eta$,
which can be estimated by the size dependence of the largest $S(\bm{q})$.
Hereafter, we choose $J$ as the unit of energy in the Blume-Capel model.

\subsection{Comparisons with the properties of the Blume-Capel model}

\begin{figure}[t]
\includegraphics[width=0.9\linewidth]{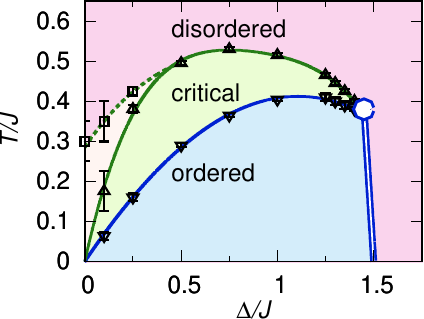}
\caption{Phase diagram of the Blume-Capel model
obtained by Monte Carlo calculations.
The first-order transition line (double line)
and the two BKT-like transition lines (single lines)
merge at the critical point (open circle).
The long-range ordered phase is characterized by the ${\uparrow}{-}{0}{-}{\downarrow}$
spin configuration.
Up- and down-pointing triangles denote
the upper and lower critical temperatures determined by the points
at which the exponents $\eta$ obtained in the largest system sizes coincide with
$\eta_{\rm low}=1/9$ and $\eta_{\rm high}=1/4$, respectively.
Squares on a dotted line
indicate the temperatures at which $\eta$ starts to
depend on system sizes.
This exponent $\eta_{\rm high}^{\rm fss}$ 
estimated similarly 
to $\eta_{\rm high}$ in Fig.~\ref{fig:exponent_eta} by finite-size
scaling
seems to exceed the expected one
$\eta_{\rm high}^{\rm clock}=1/4$
for $\Delta/J<0.5$
within the system sizes we have studied.
However, such a behavior for $\Delta < T\ln 2$,
which corresponds to $U < 0$
in Eq.~(\ref{eq:Delta_eq_halfU_plus_T}),
does not appear in the phase diagram for $U > 0$
(Fig.~\ref{fig:phase_diag}).}
\label{fig:phase_diag_BC}
\end{figure}

We compare the phase diagram of the extended Hubbard model
(Fig.~\ref{fig:phase_diag})
with that of the Blume-Capel model
(Fig.~\ref{fig:phase_diag_BC}).
The ground state shows an
${\uparrow}{-}{0}{-}{\downarrow}$-type three-sublattice magnetic
long-range order~\cite{collins1988,grigelionis1994}
for $0<\Delta/J<1.5$.
At $\Delta=0$, the ground state has quasi-long-range
order~\cite{nagai1993,lipowski1995,zeng1997,zukovic2013b}.
The first-order transition line
bifurcates at the critical end point $(\Delta_{\rm c}/J,T_{\rm c}/J)=(1.45(5),0.38(2))$
into the two boundaries of the critical phase as in the extended Hubbard
model. This result suggests that the electron spin degrees of freedom
are irrelevant for the stability of the charge BKT-like phase.
Note that in a previous study~\cite{zukovic2013a} the critical
phase was overestimated because the upper boundary was
determined by the temperature at which the specific heat takes a
maximum value, which is usually higher than the boundary of the critical phase
[see Fig.~\ref{fig:spec_heat}(b)].

Although qualitative behavior of the two phase diagrams
is similar,
the spin degrees of freedom significantly affect the phase boundary.
The first-order transition line drops nearly vertically in the
Blume-Capel model ($\delta\Delta/\delta T\simeq 0$), while it is tilted in the
extended Hubbard model at low temperatures.
The slope of the first-order transition
line in the extended Hubbard model can be estimated by
Eq.~(\ref{eq:Delta_eq_halfU_plus_T}) to be 
\begin{equation}
 \frac{\delta T}{\delta U} = -\frac{1}{2\ln 2} \simeq -0.721.
\end{equation}
This value is in agreement with the slope of first-order transition line
at low temperatures in Fig.~\ref{fig:phase_diag}. 
Note that the tilt of the phase boundary itself was also reported in the
models on a square lattice where the BKT-like phase is
absent~\cite{pawlowski2006}.

The effect of entropy becomes clearer when we apply the Clausius-Clapeyron
relation.
In the Blume-Capel model, the
Clausius-Clapeyron relation for the entropy $S_{\rm c}$ for
$T\rightarrow 0$ is given by
\begin{equation}
\frac{\delta T}{\delta \Delta}
 = \frac{\delta M}{\delta S_{\rm c}}.
\end{equation}
Since the left-hand side diverges and $\delta M$ is nonzero,
$\delta S_{\rm c}\rightarrow 0$ for $T\rightarrow 0$.
Hence, the entropies are essentially the same
for the paramagnetic and antiferromagnetic phases. On the other hand, in
the extended Hubbard model, the Clausius-Clapeyron relation for the
entropy $S_{\rm cs}$, which involves electron spin, is given by 
\begin{equation}
 \frac{\delta T}{\delta U}
 = \frac{\delta \rho_2}{\delta S_{\rm cs}}
 = -\frac{1}{2\ln 2}.
\end{equation}
Because the density of doubly occupied sites jumps from $0$ to $1/3$ across
the phase transition from the charge-uniform phase to the charge-ordered
phase, $\delta \rho_2\ne 0$, similarly to the case of the Blume-Capel model.
However, in contrast to the case of the Blume-Capel model, the
charge-uniform phase possesses larger entropy than the charge-ordered phase
because of the spin degrees of freedom ($\delta S_{\rm cs}\ne 0$), which
causes the tilt of the first-order transition line.

\section{Discussions}
\label{sec:discussions}

\subsection{Stability of the charge critical phase}

So far, we have considered the possible charge critical phase in the ideal limit.
However, the phase could be susceptible to external perturbations.
Here we discuss how the charge critical phase and signatures of it
survive in such cases.

First, the present critical phase appears at nonzero temperatures, and therefore,
external perturbations that would alter the ground state do not necessarily
destroy the critical phase.
Second, perturbations smaller than thermal fluctuations should be irrelevant
to realize the critical phase.
Even in the presence of quantum fluctuations,
the charge critical phase is expected to survive
if the transfer hopping $t$ or the spin exchange interaction $J\sim t^2/U$
is sufficiently smaller than the energy scale of temperature $T$.
Furthermore, similar arguments hold for the small perturbations, such as
weak long-range or anisotropic Coulomb interactions,
small doping,
and small bond or site randomness.
On the other hand, 
degeneracy plays an important role in the
present critical phase.
If the perturbations 
substantially destroy the degeneracy
at very low temperatures
by which
the system cannot be regarded as the effective $p$-state clock
model~\cite{jose1977} with $p\ge 5$,
the critical phase may disappear.

The situation would be much more complex
when quantum fluctuations are more dominant
than thermal fluctuations.
At very low temperatures,
it is not so obvious whether
quantum fluctuations
weaken or
stabilize the charge critical insulating phase, if any.
Moreover, the Mott transition is expected
when the transfer hopping becomes comparable to or larger than $U$ and $V$.
The nature of the Mott transition
in the Coulomb gas model with the fermionic degrees of freedom
was discussed at the level of correlated wave functions
in two dimensions in the ground state~\cite{capello2006}.
However, it is much harder to investigate a microscopic model
at low but nonzero temperatures.
Accurate determination of the phase boundaries as a function of $t/V$ and
doping concentration will be left for future study.

\subsection{Possibility of the charge critical phase in materials}

One promising candidate 
to realize the critical phase in materials
would be organic compounds which form triangular
structures~\cite{hotta2003,
mori2003,
watanabe2006,
udagawa2007,
canocortes2011,
hotta2012,
yoshimi2012,
merino2013,
tocchio2014,
mishmash2015,
gomes2016,
kaneko2017}.
Another candidate would be adatoms on semiconductor
surfaces~\cite{tosatti1974,carpinelli1996,santoro1999,hansmann2013,adler2018}.
Indeed, the $012$-type charge-ordered state is observed in Sn/Ge(111)~\cite{cortes2013}.
Both systems 
can be effectively described by
the extended Hubbard model
on a
triangular lattice,
and the Coulomb interactions, including the nonlocal one,
are estimated to be much larger than the hopping~\cite{nakamura2012,hansmann2013}.
When the effects of anisotropy and randomness are not too large
and the systems can be controlled to be at half filling,
the charge critical phase may be achievable.

It might be useful to first investigate two broad peaks
in the specific heat as a function of temperature
to detect the charge critical phase.
As we have shown in Fig.~\ref{fig:spec_heat},
the BKT-like transitions are not identical to
the points where the specific heat shows maxima.
However, these will be weak evidence of the charge critical phase.
The lower boundary of the BKT-like phase should be located 
above
the lower-temperature peak in the specific heat,
and the upper boundary of the BKT-like phase should be located
below
the higher-temperature peak in the specific heat.

After confirming the approximate transition points,
one should take the snapshots of real space charge configuration
in the intermediate-temperature region
between the charge-ordered and disordered phases.
In the charge critical phase, the number of singly occupied sites
is decreased since thermal fluctuations generate the $002$ or $022$
triangles by breaking the $012$ one
[see Fig.~\ref{fig:snapshot}(b)].

\subsection{Other BKT-like phases}

In the present paper we have investigated
the extended Hubbard model on a triangular lattice at half filling in
the atomic limit to clarify the nature of the charge BKT-like phase
caused by competing short-range Coulomb repulsion.
Similar BKT-like critical phases in the discrete degrees of freedom,
such as orbital and chirality, as well as charge,
are also expected in other lattice systems for different fillings.
For instance,
based on possible superlattice structures
in adsorbed monolayers~\cite{domany1978,domany1979},
one can consider the effective extended Hubbard model at appropriate filling,
where the ground state shows corresponding charge order.
When the given structures have degenerate configurations such that the
system can effectively be regarded as the $p$-state clock model with
$p\ge 5$ at low temperatures, the BKT-like critical phase would emerge
at finite temperatures~\cite{jose1977}. This would also hold true for
the orbital and chirality degrees of freedom. Searching for the charge,
orbital, and chirality BKT-like phases in other models is left for future
study.

\section{Summary}
\label{sec:summary}

We numerically showed that the charge BKT-like phase appears in the
intermediate-temperature regime
between the charge-ordered and disordered phases
in the extended Hubbard model on a
triangular lattice at half filling in the atomic limit. In contrast to
the conventional BKT phase in the Coulomb gas model with the logarithmic
long-range interaction, the present BKT-like critical phase is caused
only by the on-site and nearest-neighbor
Coulomb repulsion. The critical phase originates
from sixfold-degenerate charge-ordered ground states
which
generate an effective 
six-state clock model. 
The critical
phase is characterized by the algebraic decay of the charge correlation
function
accompanied
by an emergent $\rm U(1)$ symmetry
in the effective clock model at intermediate temperatures,
and the increase of the density of doubly occupied sites.
We also clarified the effect of spin degrees of freedom on the phase
diagram by comparing that in the Blume-Capel model which is an effective
model only for the charge degrees of freedom.

We believe that
the findings shown in the present paper not only serve as a good starting
point for understanding the phase diagram in two-dimensional electron systems
with geometrical frustration but also open up the possibility of searching
for novel finite-temperature critical phases that can evolve from conventional
charge-ordered phases in two-dimensional electron systems.
Experimental realization
of the charge BKT-like critical phase in materials is desired.


\acknowledgments

R.K.\ acknowledges fruitful discussions with 
Masamichi Nishino and Tsuyoshi Okubo.
The Mersenne Twister random-number generator~\cite{matsumoto1998}
was used for Monte Carlo calculations.
The numerical calculation was partly performed on
the Numerical Materials Simulator at the National Institute for Materials Science.

  
\bibliographystyle{apsrev4-1}
\bibliography{references.bib}


%

\end{document}